\documentclass[prb,showpacs,superscriptaddress,floatfix,twocolumn]{revtex4}
\usepackage{amsfonts}
\usepackage{stmaryrd}
\usepackage{bbm}
\usepackage{mathrsfs}
\usepackage{tipa}
\usepackage{amssymb}
\usepackage{txfonts}
\usepackage{graphicx}
\usepackage{dcolumn}
\usepackage{epstopdf}
\usepackage{array}

\newcommand{\PreserveBackslash}[1]{\let\temp=\\#1\let\\=\temp}
\newcolumntype{C}[1]{>{\PreserveBackslash\centering}p{#1}}
\newcolumntype{R}[1]{>{\PreserveBackslash\raggedleft}p{#1}}
\newcolumntype{L}[1]{>{\PreserveBackslash\raggedright}p{#1}}

\begin{document}

\newcommand*{\cm}{cm$^{-1}$\,}


\title{Optical spectroscopy study of the collapsed tetragonal phase of CaFe$_2$(As$_{0.935}$P$_{0.065}$)$_2$ single crystals}

\author{X. B. Wang}
\author{H. P. Wang}
\author{T. Dong}
\affiliation{Beijing National Laboratory for Condensed Matter Physics, Institute of Physics, Chinese Academy of Sciences, Beijing 100190, China}
\author{R. Y. Chen}
\affiliation{International Center for Quantum Materials, School of Physics, Peking University, Beijing 100871, China}

\author{N. L. Wang}
\email{nlwang@pku.edu.cn} \affiliation{International Center for Quantum Materials, School of Physics, Peking University, Beijing 100871, China}
\affiliation{Collaborative Innovation Center of Quantum Matter, Beijing, China}

%

\begin{abstract}

We present an optical spectroscopy study on P-doped CaFe$_2$As$_2$ which experiences a structural phase transition from tetragonal to collapsed tetragonal (cT) phase near 75 K. The measurement reveals a sudden reduction of low frequency spectral weight and emergence of a new feature near 3200 \cm (0.4 eV) in optical conductivity across the transition, indicating an abrupt reconstruction of band structure. The
appearance of new feature is related to the interband transition arising from the sinking of hole bands near $\Gamma$ point below Fermi level in the cT phase, as expected from the density function theory calculations in combination with the dynamical mean field theory. However, the reduction of Drude spectral weight is at variance with those calculations. The measurement also indicates an absence of the abnormal spectral weight transfer at high energy (near 0.5-0.7 eV) in the cT phase, suggesting a suppression of electron correlation effect.

\end{abstract}

\pacs{74.25.Gz, 74.70.Xa}

\maketitle

\section{Introduction}

Understanding the relationship between electron correlations, magnetism, and unconventional superconductivity has been a focus in the current
study of iron-based superconductors. Iron pnictides/chacolgenides are multi-orbital systems, different Fe 3d orbitals turn out to have different fillings and band widths.\cite{Lu} The electronic correlations are also found to be orbital dependent and highly influenced by Hund's coupling.\cite{Luca2,C1,C2,WGYin,C3,DFT-DMFT,DFT-DMFT2,Calderon} As thus, the orbital-dependent electronic correlation is of particular importance for understanding the normal state electronic properties and the pairing mechanism of the iron-based superconductors.

Experimentally, the "122" compound CaFe$_2$As$_2$ provides an excellent opportunity to study the correlation effect and its role on
superconductivity as the compound appears to be easily tuned into different electronic phases. At ambient pressure CaFe$_2$As$_2$ undergoes an antiferromagentic (AFM) phase transition coupled with structural distortion at 170 K,\cite{CFA} being similar to two other 122 parent compounds AFe$_2$As$_2$ (A=Ba, Sr).\cite{BFA,SrFA} Superconductivity could be induced by hole or electron doping.\cite{Na,K,Co,Rh,Saha} On the other hand, it is found that CaFe$_2$As$_2$ is rather sensitive to pressure.\cite{P1,P2,P3,cT1,P5} With hydrostatic pressure, it undergoes a structural transition from a tetragonal (T) phase to a collapsed tetragonal (cT) phase.\cite{cT1} The cT phase has the same crystal symmetry as the high-temperature T phase but with an abrupt $\sim$10\% reduction in c-axis parameters and an 2\% increase along the a-axis. The cT phase can also be stabilized at ambient pressure by chemical substitution on As site,\cite{As} Fe site,\cite{Rh} and Ca site.\cite{Ca,Sr}

The T to cT phase transition is linked with the formation of direct As-As interlayer bond and a weakening of in-plane Fe-As bonds \cite{Yildirim,cT1}. Magnetic and electronic structures are found to be strongly influenced by the transition. In the cT phase, the Fe local moment is quenched \cite{XES,NMR1}, the AFM order or magnetic fluctuation disappears \cite{neutron1,neutron2,NMR2} and the standard Fermi liquid behavior recovers.\cite{As,Rh} In addition, in sharp contrast to the paramagnetism in the T phase, intrinsic superconductivity is absent in the cT phase. Recent angle-resolved photoemission spectroscopy (ARPES) measurements show that the two electron pockets at the \emph{M} point transform into one cylinder, and the hole pockets at $\Gamma$ disappear,\cite{ARPES1,ARPES2,ARPES3} being in
agreement with the band structure calculations.\cite{cal1,cal2,As} However, in a de Haas-van Alphen effect study on CaFe$_2$P$_2$, which is a structural analogue of the cT phase of CaFe$_2$As$_2$, the hole
pockets are not completely vanished, two small hole pockets centered at Z point still exist.\cite{CFP} Very recently, two density function theory (DFT) calculations in combination with the dynamical mean field theory (DMFT) (DFT-DMFT) have been performed on both T and cT phases in an effort to
address the electron correlation and its effect on electronic properties \cite{DFT-DMFT,DFT-DMFT2}. It is found that the electron correlation becomes weaker for all Fe 3d
orbitals in cT phase. Especially, the Fe 3d$_{xy}$ orbital undergoes a change from being the most strongly renormalized orbital in the T phase to
the least renormalized orbital in the cT phase \cite{DFT-DMFT2}. The calculated optical conductivity indicates a kinetic energy gain due to the loss in Hund's rule coupling energy in the cT phase. In addition, a new feature near 0.5 eV emerges in the optical conductivity \cite{DFT-DMFT}.

It is highly desirable to investigate experimentally the charge dynamics in the cT phase and to compare with the DFT-DMFT calculations. In this work, we present a transport and an optical spectroscopy study of the collapsed tetragonal phase of
CaFe$_2$(As$_{0.935}$P$_{0.065}$)$_2$ single crystals. The isovalent substitution of As with P in CaFe$_2$As$_2$ does not introduce extra
carriers, but induce a structural transition to the cT phase near 75 K. An interfacial or filamentary superconductivity at about
30 K emerges at the same time. Our optical spectroscopy measurement reveals a sudden spectral change at low frequencies, indicating an abrupt reconstruction of band structure. A new feature appears near 0.4 eV, being roughly consistent with the DFT-DMFT calculations. However, the overall Drude-like spectral weight is lowered in the cT phase. We analyze the low-frequency data in terms of a Drude-Lorentz approach containing a narrow and a broad Drude components. On the other hand, the spectral weight transfer at high energy, commonly observed for AFe$_2$As$_2$ (A=Ba, Sr, Ca), is absent in the cT phase, which could be assigned to the weakening of the Hund's coupling energy. The sudden weakening of electron correlation appears to be correlated to the vanishing of intrinsic superconductivity in the collapse cT phase.

\section{\label{sec:level2}Experiment}

High quality of CaFe$_2$(As$_{1-x}$P$_x$)$_2$ single crystals with nominal composition of x = 0.065 were grown by the self-flux method. The
typical size was about 5$\times$5$\times$0.1 mm$^3$. The actual x values was about 0.0575$\pm$0.008 determined by energy dispersion X-rays
spectrum (EDX) on several pieces of samples in the same batch. Resistivity measurements were performed with a Quantum Design Physical Property
Measurement System (PPMS). Magnetization was measured using a Quantum Design superconducting quantum interference device (SQUID-VSM). The
optical reflectance measurements were performed on Bruker IFS 113v and 80v spectrometers in the frequency range from 50 to 35 000 \cm. An in
situ gold and aluminum overcoating technique was used to obtain the reflectivity R($\omega$). The real part of conductivity $\sigma_1(\omega)$
is obtained by the Kramers-Kronig transformation of R($\omega$). The Hagen-Rubens relation was used for low frequency extrapolation; at high
frequency side a $\omega^{-1}$ relation was used up to 300 000 \cm, above which $\omega^{-4}$ was applied.

\section{\label{sec:level2}Results and discussion}

\begin{figure}[b]
\includegraphics[clip,width=3.3 in]{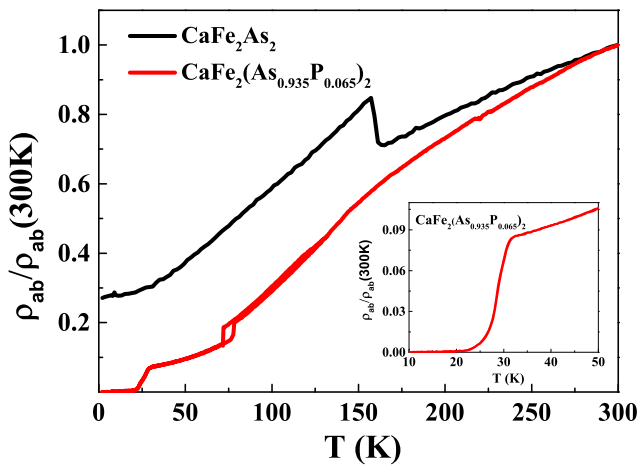}
\caption{(Color online) Temperature-dependence of normalized resistivity of CaFe$_2$As$_2$ and CaFe$_2$(As$_{0.935}$P$_{0.065}$)$_2$. The inset
plots an enlarged region for superconducting transition in CaFe$_2$(As$_{0.935}$P$_{0.065}$)$_2$.}
\end{figure}

Figure 1 shows the temperature dependence of in-plane resistivity for CaFe$_2$(As$_{1-x}$P$_x$)$_2$ single crystals at zero field. In the case
of x = 0, the parent CaFe$_2$As$_2$ shows a sharp increase in resistivity at T$_{N}$ $\approx$ 165 K, corresponding to a first-order structure
transition from the tetragonal phase to the orthorhombic AFM phase. By substituting isovalent P for As, the SDW state is suppressed and an
abrupt decrease of resistivity occurs at about T$_{cT}$ $\approx$ 75 K upon cooling. The large thermal hysteresis about 10 K suggest the
first-order nature of the transition. With a further decreasing in temperature, the resistivity shows a T$^{2}$ dependence, which indicates the
recovery of Fermi liquid behavior, consistent with the previous reports.\cite{As,Rh} Notably, there is another obvious
down-turn at about 32 K, and the resistivity reaches zero at about 20 K, as shown in the inset of Fig. 1. This is very similar to the case of
rare-earth-doped CaFe$_2$As$_2$, where the superconductivity is considered as an interfacial or filamentary effect. This is very interesting
because it was suggested that isovalent substitution of P for As in CaFe$_2$As$_2$ couldn't induce superconductivity before.\cite{As}

\begin{figure}[t]
\includegraphics[clip,width=3.3in]{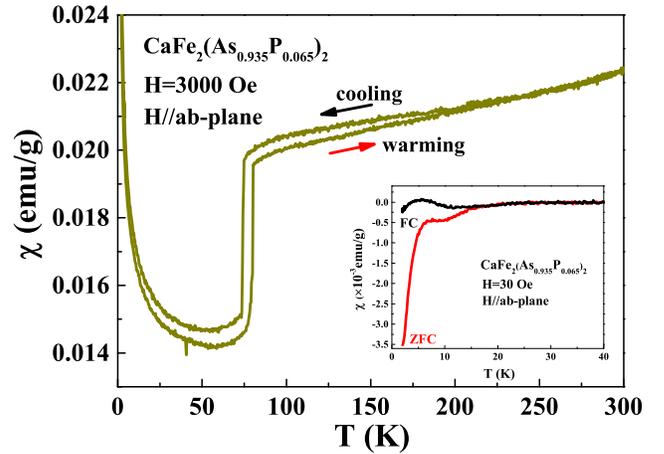}
\caption{(Color online) Temperature-dependence of magnetic susceptibility of CaFe$_2$(As$_{0.935}$P$_{0.065}$)$_2$ with \textbf{H}
\emph{$\parallel$}$ \emph{ab}$-plane at 3000 Oe. The inset shows the zero-field-cooled and field-cooled data at 30 Oe.}
\end{figure}

\begin{figure*}[t]
\includegraphics[clip,width=7.1in]{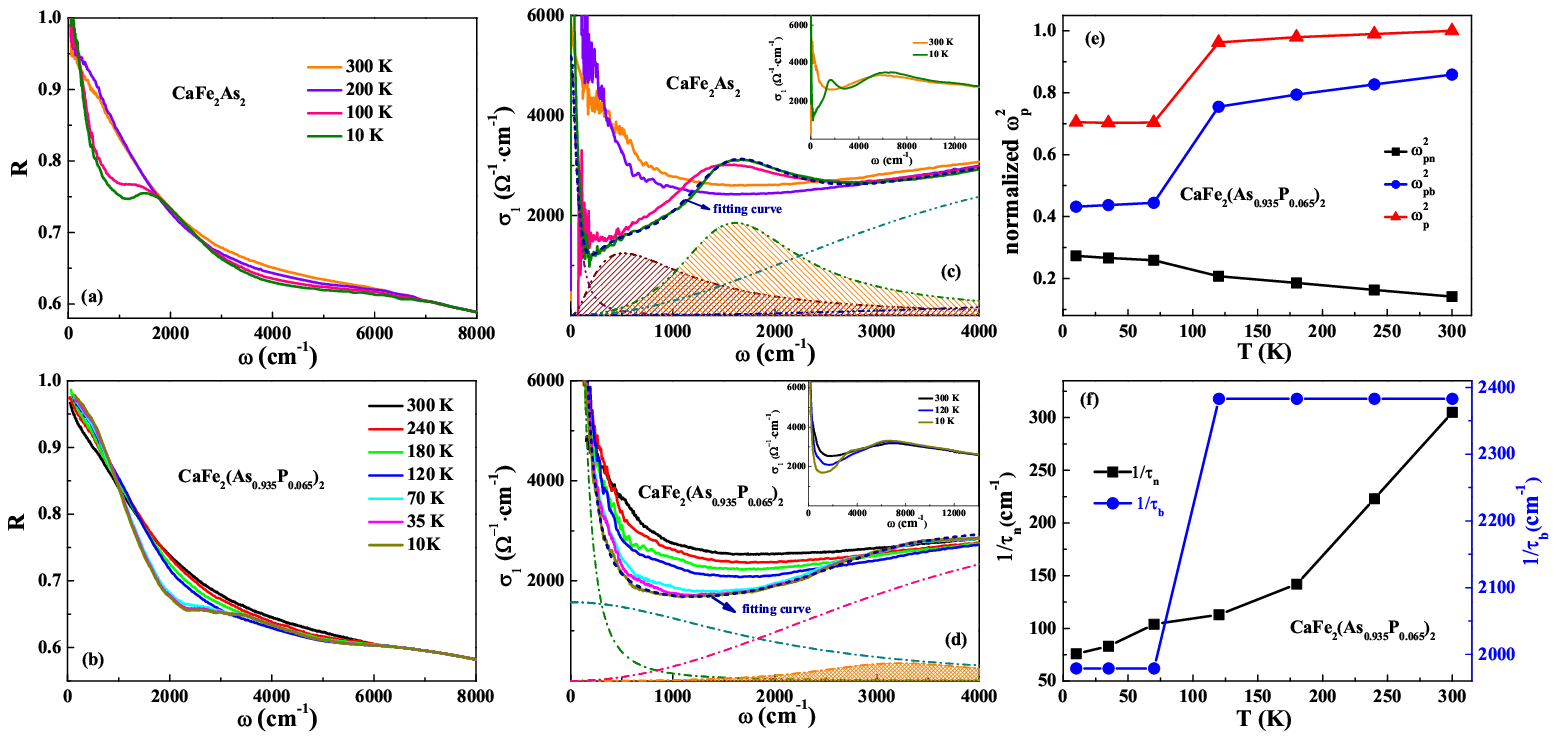}
\caption{(Color online). Left panel: {R($\omega$)} for (a) CaFe$_2$As$_2$ and (b) CaFe$_2$(As$_{0.935}$P$_{0.065}$)$_2$ below 8000 \cm. Middle
panel: {$\sigma$$_1$($\omega$)} for (c) CaFe$_2$As$_2$ and (d) CaFe$_2$(As$_{0.935}$P$_{0.065}$)$_2$ blow 4000 \cm. The Drude-Lorentz fit for T
= 10 K are shown at the bottom. Inset: the optical conductivity over a much wider frequency range. Right panel: Temperature dependence of
$\omega_p^2$ normalized to the value of 300 K and scattering rate1/$\tau$ of the two Drude terms of CaFe$_2$(As$_{0.935}$P$_{0.065}$)$_2$.}
\end{figure*}

The temperature dependence of dc susceptibility of CaFe$_2$(As$_{0.935}$P$_{0.065}$)$_2$ single crystals in an applied magnetic filed of H = 0.3 T
aligned in ab-plane is shown in Fig. 2. The susceptibility decreases slightly with cooling temperatures, but drops abruptly below T$_{cT}$. The strong Curie-Weiss-like tail at lower temperature is not intrinsic and can be attributed to the presence of paramagnetic impurities or defects, since Knight shift measurements on compounds showing similar Curie-Weiss behavior in the cT phase did not reveal any upturn.\cite{NMR2} The huge
thermal hysteresis and sudden decrease of susceptibility provides further evidence of first-order transition in this compound. At 30 Oe, as
shown the inset of Fig. 2, a diamagnetic signal of superconducting transition at about 15 K, and a rapid drop of the dc susceptibility at 7 k
are observed. This is very similar to Pr-doped CaFe$_2$As$_2$, suggesting the possible existence of two superconducting phases.\cite{Pr} However
as shown in the Fig. 2, the superconductivity is killed at a small magnetic filed (H = 0.3 T), consistent with interfacial or filamentary
superconductivity scenario. Both the resistivity and magnetic properties suggest that the SDW state is supressed and the system exhibits a
first-order structural phase transition from tetragonal to collapsed-tetragonal phase in CaFe$_2$(As$_{0.935}$P$_{0.065}$)$_2$ single crystals.

Figure 3(a) and (b) show the reflectance spectra of the CaFe$_2$(As$_{1-x}$P$_x$)$_2$ crystals (x = 0 and 0.065) at different temperatures. The
reflectance exhibits a metallic response in both frequency and temperature dependence, and the most obvious feature is the suppression of
R($\omega$) in both compounds at low temperatures. As a consequence, relatively sharp reflectance edges appear at low frequency, which suggests a
reduction of the carrier scattering rate. In the parent CaFe$_2$As$_2$, whose behavior is very similar to that of the other two members of
AFe$_2$As$_2$ system (\emph{A} = Ba, Sr),\cite{AFeAs} the significant decrease in energy scale of the reflectance edge implies a considerable reduction of carrier density. While in the P-doped CaFe$_2$As$_2$, there is a sudden suppression of the reflectance between 1000 \cm to 3500 \cm across the T to cT phase
transition. As a result, the spectra were separated into two groups above and below T$_{cT}$, suggesting an abrupt band reconstruction as well. Nevertheless, the spectral change is distinctly different from the AFM phase for the parent compound.

The middle panels of Fig. 3 show the conductivity spectra $\sigma$$_1$($\omega$) below 4000 \cm. In the parent compound, the spectra is
severely suppressed at low frequencies, only a very sharp and narrow Drude term is left, and a double-peak feature appears blow T$_N$ which are
known as the SDW gaps \cite{AFeAs} as shown in the Fig. 3 (c). While in the P-doped compound, the optical conductivity is suppressed suddenly below T$_{cT}$
at much higher and wider frequency range. A peak-like structure
appears near 3200 \cm (0.4 eV), which we shall discuss later. The data are in agreement with a recent DFT-DFMT calculation of the collapsed tetragonal
phase of CaFe$_2$As$_2$.\cite{DFT-DMFT}

To make a quantitative analysis of the temperature evolution of the different part of electronic excitations, we decompose the optical
conductivity using a simple Drude-lorentz model:\cite{AFeAs}
\begin{equation}
\epsilon(\omega)=\epsilon_\infty-\sum_{i}{{\omega_{p,i}^2}\over{\omega_{i}^2+i\omega/\tau_{i}}}+\sum_{i}{{\Omega_i^2}\over{\omega_i^2-\omega^2-i\omega/\tau_i}}.
\label{chik}
\end{equation}
where $\epsilon_\infty$ is the dielectric constant at high energy, and the middle and last terms are the Drude and Lorentz components,
respectively. The Drude components represent the contribution from conduction electrons, while the Lorentz components describe the
interband transitions.

We are mainly concerned about the evolution of the low-energy optical conductivity. Unlike the parent CaFe$_2$As$_2$, the low-energy optical
conductivity of CaFe$_2$(As$_{0.935}$P$_{0.065}$)$_2$ can fit very well by two Drude terms, a narrow one and a broad one, at all temperatures as
shown in Fig. 3(d). The two Drude components may provide information about different types of carriers in this compound.\cite{Homes} The
temperature dependence of normalized plasma frequency $\omega^2_p$ and scattering rate 1/$\tau$ are shown in Fig. 3 (e) and (f). The subscripts
{\sl{n}} and {\sl{b}} stand for the narrow and broad Drude terms, respectively.

The plasma frequency $\omega^2_{p,b}$ undergoes a discontinuous decrease at T$_{cT}$, however $\omega^2_{p,n}$ shows an opposite trend. Nevertheless, the total low energy plasma frequency $\omega^2_p$ = $\omega^2_{p,n}$+$\omega^2_{p,b}$ decreases by
about 20\% in the cT phase, indicating an overall decrease of n/m$^\ast$. Very recently, the angle resolved photoemission spectroscopy (ARPES) investigations on the cT phase of CaFe$_2$As$_2$ single crystals suggest that the hole pockets near $\Gamma$ point sink below the Fermi level while the electron band moves to a higher binding energy, resulting in a decrease of the band effective mass of electron pocket.\cite{ARPES1,ARPES2,ARPES3} Evidently, if we associate the narrow Drude component with the electron pocket and the broad one with the hole pocket, we can attribute the increase of $\omega^2_{p,n}$ to the reduction of band mass of electron pocket and the decrease of $\omega^2_{p,b}$ to the partial removal of hole pocket (mainly close to $\Gamma$ point). In an earlier study on various 3d transition metal based 122 compounds, Cheng et al. found that the moderate 3d-electron correlation is responsible for the incoherent term (the broad Drude
term here) in the iron pnictides.\cite{BCheng} Thus the decrease of $\omega$$^2$$_{p,b}$ suggests a suppression of correlation effect, which is consistent with recent NMR and theoretical study.\cite{NMR2,DFT-DMFT} It deserves to remark that, although the effective mass from electron Fermi surface is indeed reduced in the cT phase,\cite{ARPES2,DFT-DMFT} the overall plasma frequency $\omega^2_p$ is still reduced due to the large reduction of hole carrier
density. Since the kinetic energy of the electrons is proportional to the plasma frequency, the measurement indicates that the overall kinetic energy of charge carriers decreases in the cT phase. The result is different from the DFT-DMFT calculations that suggests an increase of the low frequency spectral weight in the cT phase. We notice that, in the above-mentioned calculations, the optical kinetic energy was estimated by summarizing the conductivity spectrum up to a certain cutoff frequency, and the same cutoff frequency was used for both T and cT phases. In fact, the Drude spectral weight is shifted to lower frequency in cT phase in comparison with T phase. Therefore, different cutoff frequencies should be used in summarizing the Drude components of conductivity for the two phases.
Further work is needed to address this issue.

Figure 3 (f) shows the T dependence of the scattering rate of the two Drude components. The scattering rate of the narrow Drude term 1/$\tau$$_n$ decreases upon cooling and there is a downturn below T$_{cT}$, while the 1/$\tau$$_b$ decreases suddenly below T$_{cT}$. In the optimally doped K- and P-doped BaFe$_2$As$_2$, 1/$\tau$$_n$ decreases linearly with temperature and 1/$\tau$$_b$ keeps constant in the whole temperature range
above T$_c$ \cite{K-Ba122}. Then, the sudden changes of 1/$\tau$$_n$ and 1/$\tau$$_b$ in CaFe$_2$(As$_{0.935}$P$_{0.065}$)$_2$ upon entering the cT phase may be related to the suppression of spin fluctuation or the reduction of scattering channel caused by the loss of hole pockets near $\Gamma$ point.

\begin{figure}
\includegraphics[clip,width=3.3in]{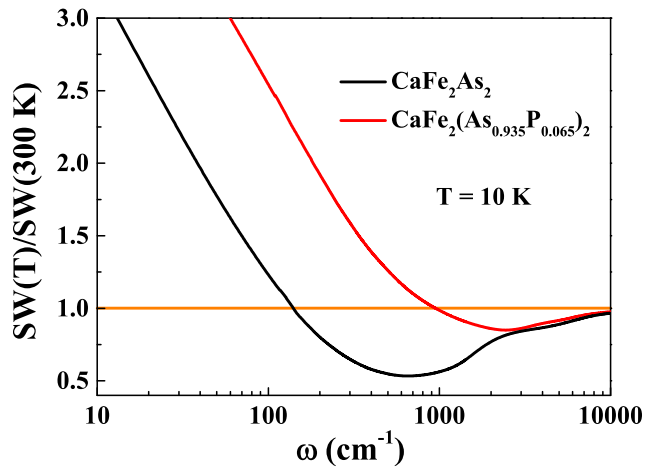}
\caption{(Color online). Ratio of the integrated spectral weight as a function of $\omega$ at 10 K in CaFe$_2$As$_2$ and
CaFe$_2$(As$_{0.935}$P$_{0.065}$)$_2$. The low-temperature integrated spectral weight is normalized to the room-temperature data.}
\end{figure}

To further elaborate on the spectral evolution, we plot the frequency-dependent integrated spectral weight of CaFe$_2$(As$_{1-x}$P$_x$)$_2$ (x =
0 and 0.065) at 10 K, as shown in Fig. 4, and the low-temperature integrated spectral weight has been normalized to the integrated spectral weight at room
temperature. For the parent CaFe$_2$As$_2$, the strong dip blow 1000 \cm represents the formation of the SDW gaps, and the small turn point
around 4000 \cm is attributed to unconventional spectral weight transformation due to Hund's coupling.\cite{SW} In the P-doped compound, there
is no SDW gap that develops at low temperature, and the low-\emph{$\omega$} spectral weight change is induced simply by the narrowing of Drude component. Additionally, a new peak emerges starting from 3000 \cm, leading to a gradual recovery of the spectral weight at very high energy. As we have mentioned, the hole bands at $\Gamma$ point, which cross the Fermi level above T$_{cT}$, suddenly sink
below the Fermi level in the cT phase, as a consequences, the original intraband transition in the T phase would vanish in the cT phase and an extra interband transition around
$\Gamma$ point would appear. The structure is also evident in the DFT-DMFT calculations but with the peak centered at slightly higher energy 0.5 eV.\cite{DFT-DMFT}

Finally, we'd like to comment shortly on the spectra at higher energies. It is well established that, for the iron-pnictides/chacogenides, there exist a temperature-induced spectral weight transfer at high energies. With decreasing temperature, the low-frequency spectral weight is
transferred to the high-energy region (usually above 0.5$\sim$0.7 eV) \cite{AFeAs,SW,Schafgans,RHYuan}. Such spectral weight transfer was ascribed to the electron correlation effect, in particular, to the Hund's
coupling effect between itinerant Fe 3d electrons and localized
Fe 3d electrons in different orbitals  \cite{SW,Schafgans}. It represents the redistribution of the spectral weight
between different 3d bands \cite{SW,RHYuan}. In the present P-doped sample, this temperature dependent spectral weight transfer at high energy is not visible in the cT phase, as shown in the inset of middle
pinel of Fig. 3. It provides further evidence that the electron correlation is suppressed in the cT phase. The result is also consistent with the DFT-DMFT calculations which indicates a loss of Hund's coupling energy. The suppression of the electron correlations, including the Hund's coupling, is apparently linked with the loss of Fe magnetic moment and disappearance of superconductivity in the cT phase.

\section{\label{sec:level2}Summary}

In summary, a systematic investigation of resistivity, susceptibility, and optical spectroscopy is performed on the collapsed tetragonal phase
of CaFe$_2$(As$_{0.935}$P$_{0.065}$)$_2$ single crystals. The isovalent substitution of P for As suppresses the SDW state and results in a
first-order structural phase transition from tetragonal to collapsed-tetragonal phase. It also induces an interfacial or filamentary superconductivity near 30 K. Related to the change in the topology of the Fermi surface, the optical spectroscopy shows a sudden change. The low-energy
optical spectrum can be decomposed into two components, a broad and a narrow Drude in the whole temperature range. The narrow Drude term
increases its spectral weight in the cT phase which can be attributed to the decrease of the band effective mass of electron pocket, whereas the weight of broad Drude term decreases abruptly being likely due to the vanish of the hole pockets near $\Gamma$ point, which also leads to an additional mid-infrared peak in the cT phase. Furthermore, our data reveal that the electron
correction is suppressed, which is correlated with the loss of Fe magnetic moment and quenching of superconductivity in the cT phase.

\begin{acknowledgments}

We acknowledge useful discussions with Subhasish Mandal. This work is supported by the National Science Foundation of China (11120101003, 11327806), and the 973 project of the Ministry of Science and
Technology of China (2011CB921701, 2012CB821403).

\end{acknowledgments}

\end{document}